\documentclass[seceq,supplement]{ptptex}

\newcommand{\be}{\begin{eqnarray}}
\newcommand{\ee}{\end{eqnarray}}

\def\dd{{\rm d}}

\title{
String Theory on Thin Semiconductors~\footnote{Invited talk given at the conference ``30 Years of Mathematical Methods in High Energy Physics'',
held at RIMS, Kyoto University, March 2008. To appear in the Proceedings.}%
}

\subtitle{Holographic Realization of Fermi Points and Surfaces}    

\author{
Soo-Jong \textsc{Rey}%
}

\inst{
School of Physics and Astronomy \\
Center for Theoretical Physics \\
Seoul National University, Seoul 151-747 KOREA
}

\abst{
In this talk, I make a novel contact between string theory and degenerate fermion dynamics in thin semiconductors. Utilizing AdS/CFT correspondence in string theory and tunability of coupling parameters
in condensed matter systems, I focus on the possibilities testing string theory from tabletop experiments.
I first discuss the observation that stability of Fermi surface is classifiable according to K-theory.
I then elaborate two concrete realization of Fermi surfaces of zero and two dimensions. Both are realized by complex of D3-branes and D7-branes of relative codimension 6 and 4, respectively. The setup with Fermi point models gauge dynamics of multiply stacked graphenes at half-filling. I show that string theory predicts dynamical generation of mass gap and metal-insulator quantum phase transition at zero temperature. I emphasize that conformally invariant gauge theory dynamics of the setup plays a crucial role, leading to novel conformal phase transition. The setup with Fermi surface is based on charged black hole and models relativistic Fermi liquid. We find positive evidence for this identification from both equilibrium thermodynamics at or near zero temperature and out-of-equilibrium linear response and transport properties. I argue that fluctuation of black hole horizon provides holographic realization of the Landau's Fermi liquid theory.}

\begin{document}

\maketitle

\section{Introduction}
Quantum critical phenomena refer to a criticality associated with nonanalytic rearrangement of quantum ground-state (near or at zero temperature) as the Hamiltonian is continuously perturbed \cite{QPT}. The simplest of this sort already takes place in one-particle quantum mechanics. Upon tuning coupling parameters $\{g\}$, ground-state may undergo level crossing at a critical couplings, $\{ g_{\rm cr} \}$. If the level crossing is robust against perturbation, the ground-state energy $E_0 (\{g\})$ becomes nonanalytic across $\{g\}= \{g_{\rm cr}\}$. If this happens, we say that the quantum system has undergone 'quantum phase transition'. Correlation length, time scales and mass gaps of the ground-state may also change nonanalytically across $\{g\}= \{g_{\rm cr} \}$. In quantum field theories of diverse spacetime dimensions, such quantum phase transitions are abundant at zero temperature, often in conjunction with classical phase transitions at finite temperature. Typically, quantum phase transitions involve strong correlations and ${\cal O}(1)$ to strong couplings.

Quantum critical phenomena are expected richer and more complex in string theory since backgrounds are characterized by a variety of effective couplings and parameters such as fluxes, charges, moduli, etc. Central to all these complexity are D-branes and Ramond-Ramond fields sourced by them \cite{polchinski}. In particular, via AdS$_{D+1}$ /CFT$_D$ (Anti-de Sitter / Conformal Field Theory) correspondence \cite{ads/cft}, second-order quantum phase transition in CFT may correspond to some sort of second-order classical phase transition in AdS string theory. The latter, however, is not the more familiar thermal transition. Rather, as boundary condition to bulk fields in AdS string theory is varied, the holography asserts that a nonanalytic transition ought to take place in the bulk.
In this talk, I would like to present a particular string theory setup that exhibits a quantum critical phenomena associated with Fermi surface of degenerate fermions \cite{mytalk}. Not only does it bear intrinsic interests, my aim is to demonstrate that physics of Fermi surface is rich and futile in the context of string theory.

I first recapitulate an argument that K-theory or equivariant K-theory \cite{Ktheory} are relevant for classifying {\sl stable} Fermi nodal points. The description brings out a very interesting viewpoint that puts Fermi nodal points in Landau's Fermi liquid theory reminiscent of D-branes in string theory. As such, I find that stable Fermi surfaces obeys Bott periodicity. In (3+1)-dimensional spacetime, the stable Fermi nodal points form
a Fermi point or a Fermi surface. A natural question is whether these stable Fermi nodes can be realized within string theory. In this talk, with relatively simple setups, I argue that it is indeed possible.

As the first illustration, I present a string theory setup in which the Fermi surface is zero-dimensional. I shall analyze quantum dynamics that underlie the transition. Our proposed setup is extremely simple: $N_c$ stack of D3-branes intersecting with $N_f$ stack of D7-branes along two spatial directions. What makes this setup unique is that a suitable low-energy dynamics of these D-branes may be realizable by multi-stack of graphene --- monatomic layer of graphite hexagonal lattice --- at room temperature. Therefore, it opens an exciting opportunity to test string theory dynamics at nanotechnology laboratories. I then explain a string theory setup in which the Fermi surface is zero-dimensional.  It is a complex of D3-branes and D7-branes intersecting mutually on (2+1)-dimensional spacetime. The central feature of this D-brane complex is that it breaks supersymmetry completely and that it has an overall transverse dimension. I also explain that in string theory the proposed setup is the only possibility exhibiting both features. The
setup is parametrized by $N_c, N_f, \alpha', g_s$. I take the decoupling limit $\alpha' \rightarrow 0, N_c \rightarrow \infty$ while holding $N_f$ and the `t Hooft coupling $\lambda \equiv N_c g_s$ finite. In section 3, I analyze weak `t Hooft coupling regime. I show that low-energy dynamics is
described by (3+1)-dimensional conformal gauge theory interacting with (2+1)-dimensional conformal defects.

As the second illustration, I present a string theory setup in which the Fermi surface is two-dimensional.
Again, the setup involves $N_c$ D3-branes. Optionally, $N_f$ D7-branes may be introduced. The present setup differs from the above one in that D3-branes are spinning in transverse space. At large $N$ and strong `t Hooft coupling limit, the spacetime geometry involves $AdS_5$ but with nontrivial gauge field strengths turned on and turns into a charged black hole. I show that black hole thermodynamics at both zero and nonzero temperature match perfectly with statistical mechanics of noninteracting relativistic fermions at finite density. I further show that fluctuation dynamics of the charged black hole match as well with out-of-equilibrium linear response of Landau's Fermi liquid.

\section{K-theory Classification of Fermi Surfaces}
Consider a Fermi liquid in (d+1)-dimensional spacetime. I expect that ground-state is characterized
by a generalized Fermi surface: a hypersurface in $d$-dimensional momentum space. Stable Fermi surfaces
are classified by K-theory for real fermions or equivariant K-theory for complex fermions \cite{horava}.

Consider degenerate fermions in an ambient $(d+1)$-dimensional space-time. In condensed matter, spatial rotational invariance is broken by crystal fields. Near each nodal Fermi point, the rotational invariance is restored and even enhanced to an effective Lorentz invariance in which Fermi velocity $v_F$ plays the role of the speed of light $c$ in Einstein's special relativity. Therefore, at low-energy near the Fermi energy, fermion excitations transform in the spinor representation of the Lorentz group SO(d,1) and in the fundamental representation of an internal symmetry group $G_F$. As such, without loss of generality, I can take $\psi_{A\alpha}(x)$ as $(d+1)$-dimensional relativistic fermions. Turning on finite chemical potential $\mu_A$ for each components, real time dynamics of the fermion is described microscopically
by
\be
I_{\rm low-energy} = \int \dd^{d+1} x \, \overline{\psi}_{A\alpha} \, {\delta^A}_B \otimes {(i \Gamma^a e_a^m \partial_m - \mu_A \Gamma^0 )^\alpha}_\beta \, \psi^{A\beta} + \cdots \, ,
\ee
Here, the ellipses denote interaction of the fermions with other excitations present at low-energy and $e_a^m = (1, v_F, \cdots, v_F)$ denotes `vielbein' of the $(d+1)$-dimensional space-time. Often, the massless fermions arise from different nodal points, so $G_F$ is in general mixed with spatial symmetry group and is broken softly by unequal chemical potentials $\mu_A$. I shall consider a simplifying situation that $G_F$ and spacetime symmetry group are decoupled and that the chemical potentials are all equal, $\mu_1 = \mu_2 = \cdots \equiv \mu$. The on-shell configuration is described by
\be
(\Gamma^0 \Gamma \cdot {\bf p} - \mu ) \psi_\pm({\bf p}) = \epsilon_\pm({\bf p}) \psi_\pm ({\bf p}).
\ee

Consider the fermion propagator ${S^{A\alpha}}_{B\beta} (k, \omega) = \langle \psi^{A\alpha} (k, \omega) \psi^\dagger_{B\beta} (-k) \rangle$ in the presence of the chemical potential $\mu$. Denoting the energy and momentum measured from the Fermi energy and momentum as $(\omega, {\bf k})$, the two-point function at low energy (near the Fermi nodal points) is given by
\be
{(S^{-1})_{A\alpha}}^{B\beta} = {\delta_{A\alpha}}^{B\beta}(\omega \mathbb{I} - {\bf v}_F \cdot {\bf k}) + {\Pi_{A\alpha}}^{B\beta} (\omega, {\bf k}) \, .
\label{spectralfunction}\ee
Here, the self-energy $\Pi(\omega, {\bf k})$ refers to deformations or perturbation of the otherwise free Fermi surface. The first term shows manifest spin-flavor symmetry of excitations near each nodal point. In general, through the self-energy, the two-point function mixes gapless fermion modes at different nodes (flavors) and hence defines a linear map from $\mathbb{R}^D$ for Majorana fermions or $\mathbb{C}^D$ for Dirac fermions to $(d+1)$-dimensional $(\omega, {\bf k})$ space, where $D \equiv {\rm dim} G_F \cdot 2^{[d/2]}$. This linear map is a manifestation of flavor-spin symmetry (as demonstrated by the first term (\ref{spectralfunction})) for gapless fermion excitations near each Fermi nodal point.

The Fermi surface refers to the set of Fermi nodal points. To focus on low-energy excitations, I consider taking renormalization group scaling toward the Fermi surface. Incidentally, the same scaling arises for non-relativistic fermions except that the Fermi velocity is less than the light velocity in this case.
Assuming that the self-energy is a smooth function of $(\omega, {\bf k})$, the Fermi surface is located at the zero eigenvalue of the proper spectral function $S^{-1}(\omega, {\bf k})$. For non-interacting fermions,  $\Pi (\omega, {\bf k}) = 0$.

Stability of the Fermi surface depends on whether there exists (marginally) relevant interactions. If exists, the interactions would drive the fermion excitations into massive modes. If the interactions are anisotropic, then it is also possible to drive the fermion excitations into massive modes only for certain directions. Viewing gapless and gapped fermion excitations as counterparts of open and closed string modes, respectively, the Fermi surface is a direct analog of the D-branes in string theory.

Whether a given nodal point is stable or not depends on details of dynamics. In general, the nodal point may be perturbed by strong interactions and develops an energy gap. On the other hand, the nodal point may be rigid by topological rigidity and does not develop an energy gap. As such, stability of Fermi surface is ensured if topological rigidity is present. To classify the topology, assume that nodal points form a $(d-p)$-dimensional sub-manifold $\Sigma_\parallel$ defined by $\omega = 0$ and ${\bf k}_\perp = (k^1, \cdots, k^p) = 0$  in the ambient $(d+1)$-dimensional space of $(\omega, {\bf k})$. As such, it is parametrized by ${\bf k}_\parallel = (k^{p+1}, \cdots, k^d)$. The topological stability is ensured if the nodal points, viz. zeros of inverse spectral function, are rigid under variation of $(p+1)$-dimensional space of transverse momentum ${\bf k}_\perp$. This defines the $p$-th homotopy map from a perturbation of nonzero $(\omega, {\bf k}_\perp)$ to real- or complex-valued $(D \times D)$  matrix of inverse spectral function:
\be
\Pi_{p}: \qquad \mathbb{S}^{p} \longrightarrow \mbox{GL}(D, \mathbb{R}) \quad \mbox{or} \quad
\mbox{GL}(D, \mathbb{C}).
\ee
The homotopy map belongs to the so-called stable regime, $2D> {p}$, for all situations since the regime covers $d \ge 2$ for $N_F$ arbitrary and $N_F \ge 2$ for $d \ge 1$. As is well-known, homotopy map in the stable regime defines the K-theory over the perturbation space $(\omega, {\bf k}_\perp)$:
\be
KR(\mathbb{R}^{p+1}) = \Pi_{p} (\mbox{GL}(D, \mathbb{R})) \qquad
\mbox{or} \qquad K(\mathbb{R}^{p+1}) = \Pi_{p} (\mbox{GL}(D, \mathbb{C})).
\ee
But then, the Bott periodicity asserts that $K(\mathbb{R}^{\rm even}) = \mathbb{Z}, K(\mathbb{R}^{\rm odd}) = 0$. In the former case, the cohomology theory defines topological rigidity and the nodal points are stable. This means that Fermi surfaces of even codimensions are stable while Fermi surface of odd codimensions are unstable. Therefore, stable Fermi surfaces are direct analogs of D-branes in Type IIB string theory. Is it also possible to identify counterpart of D-branes in Type IIA string theory? The answer is affirmatively yes, but with one caveat. In the above discussion, by codimensions, I referred to the spacetime that elementary fermions were originally extended. It could be that elementary fermions are already localized compared to other (such as Coulomb or retarded gauge potential) interactions. If elementary fermions were localized on "impurity" surface of one dimension less than embedding spacetime, then the stable Fermi surfaces have odd codimension. I see that Fermi surface of 'impurity' fermions are direct analog of D-branes in Type IIA string theory. On the other hand, for real fermions, nontrivial element of KR$(\mathbb{R}^p)$ is $\mathbb{Z}$ for $p=2,6$, $\mathbb{Z}_2$ for $p=3,4$, all with periodicity $8$. The $\mathbb{Z}_2$-valued Fermi surface is novel and very interseting. It would be extremely interesting to identify within string theory setups.

For the stable Fermi surface, gapless fermion excitations are described by fluctuation of the Fermi surface. Evidently, it depends only on momentum transverse to the Fermi surface, viz. ${\bf k}_\perp = (k^1, \cdots, k^p)$. Once again, the K-theory dictates universality of these fluctuation spectrum. By Atiyah-Bott-Shapiro theorem, hydrodynamic fermion excitations, which I denote as $\chi_A(\omega, {\bf k})$, at the Fermi surface is described by
\be
S_{\rm fluctuation} = \int \dd \omega \dd {\bf k}_\parallel \dd {\bf k}_\perp
\overline{\chi}_A(\omega, {\bf k})  {[\Gamma^a \tilde{e}_a^m {k}_{\perp,m} \hskip-0.75cm / \hskip0.55cm]^A}_B \chi^B (\omega, {\bf k}) + \cdots \, .
\ee
Here, the fermion spans a vector in the real or complex dimension of spin-flavor symmetry group and $\tilde{e}_a^m$ are Fermi velocity matrix when restricted to the hydrodynamic excitations. This shows that low-energy excitation of stable Fermi nodal points exhibits universally relativistic dispersion relation.
What cannot be determined from K-theory consideration alone is details of the dynamics such as Fermi velocity and interactions.

\section{Holography of Fermi Point}
I now present realization of Fermi surface of codimension $d$ --- Fermi point --- within string theory.
For reasons that will become clearer, I shall refer to the setup as "holographic graphene multilayer".
The setup is actually quite interesting since physics underlying the setup involve a variety of
phenomena such as dynamical mass generation \cite{fermionmass}, multi-flavor symmetry \cite{vafawitten},\cite{flavorsymmetrybreaking}, parity symmetry breaking \cite{vafawitten}, induced three-dimensional Chern-Simons action \cite{redlich}, supercritical instability \cite{supercritical} etc.

Graphene \cite{graphene} refers to monolayer of carbon crystal, forming hexagonal lattice. Recently, the graphene monolayer was obtained successfully with ridiculously elementary fabrication technology using pencil and scotch tape. Energy band of ideal graphene has the structure that valence and conduction bands meet at two nodal points. At half filling, the valence band is filled (realizing the Dirac vacuum) and forms a gapless semi-metal or semi-conductor \cite{gapless}. Low-energy excitations exhibit energy independent dispersion, $\epsilon({\bf p}) = v_F |{\bf p}|$. Therefore, at low-energy, graphene exhibits Lorentz invariance except the absolute velocity is now set by the Fermi velocity $v_F \simeq c/300$. The two nodal points are symmetric, so low-energy excitations are described by a pair of two-component fermions, equivalently, a four-component Dirac fermion.

Multiple stack of graphene layers can also be fabricated between a bulk substrate. Therefore, electrodynamics of half-filled graphene multi-layers forms a very interesting system with the features (1) $N_F$ flavors of 4-component Dirac fermions are localized on (2+1)-dimensional spacetime, where $N_F$ is the number of graphene multi-layers, (2) the electromagnetic interaction is mediated through (3+1)-dimensional spacetime, (3) the effective fine structure constant can be artificially tuned by varying dielectric constant $\epsilon$ of the substrate, $\alpha_{\rm eff} \simeq \alpha_e / \epsilon v_F$. Consequently, multi-stacked graphenes
are characterized by three parameters: $N_F$, $\alpha_{\rm eff}$ and temperature $T$.

I shall now show that, utilizing D-branes and AdS/CFT correspondence, gauge dynamics of graphene multi-layers can be modeled in string theory. This opens an exciting possibility that the string theory can be tested experimentally at laboratory --- even more interestingly, with relatively cheap cost and small-scale of requisite experimental setup, one can envisage that the test is achievable just around the corner in coming years!

I begin with the following D-brane setup. To model the $(3+1)$-dimensional electromagnetic interactions, I introduce $N_3$ D3-branes oriented along $(123)$ directions. To model the $(2+1)$-dimensional graphenes, I also introduce $N_F$ D7-branes along $(1245678)$ directions. Relative codimension between D3-branes and D7-branes is six, so the configuration breaks supersymmetry completely. Both D-branes have overall transverse space along $(9)$-direction. At weak coupling and low energy, $g_s, \, \ell_s \rightarrow 0$, excitations comprises of massless states of D3-D3 and D3-D7 open strings. The former forms (3+1)-dimensional ${\cal N}=4$ super Yang-Mills theory, while the latter forms $N_F$ sets of massless 4-component Dirac fermions arising from Ramond-Ramond sector.

To understand graphene electrodynamics at weak coupling, for simplicity, I shall study the U(1) gauge group. The ${\cal N}=4$ supermultiplet involves $(A_m, \psi^A, X_a)$ with $m=0, \cdots,3, A = 1, \cdots, 4, a=1, \cdots, 6$. The photinos and scalars are neutral, so the theory is non-interacting. I consider coupling a
four-current $J^m$ (provided by nonzero codimensional charge-current such as the endpoint of fundamental string on a D3-brane (codimension-2) or the fermions localized on D7-brane defect (codimension-1)). The interacting part of the theory reads, after gauge-fixing,
\be I = \int_{\mathbb{R}^4} \Big[ {1 \over 4 g^2}
F_{MN}^2 + {1 \over 2 g^2 \xi} (\partial^M A_M)^2 + \sum_{a=1}^{N_F}
(\overline{\psi}^a (i \partial \hskip-0.22cm / + A \hskip-0.22cm / ) \psi^a  \delta (x^3) \Big], \ee
In the last term, I emphasize that gapless fermions are localized on the defect $\mathbb{R}^{2,1}$ inside $\mathbb{R}^{3,1}$. I also put the defect planar transverse to $x^3$ direction.

Dynamics of D3-branes, described by ${\cal N}=4$ super Yang-Mills theory, is SO(4,2) conformally invariant.
Dynamics at the intersection with D7-branes, described by U$(2N_f)$ gapless fermions, preserves conformal subgroup SO(3,2) of SO(4,2). Therefore, the above setup is described by (3+1)-dimensional (super)conformal field theory coupled to (2+1)-dimensional conformal field theory. Equivalently, I can integrating out (3+1)-dimensional conformal field theory and obtain a nonlocal (2+1)-dimensional nonlocal conformal field theory
described by the action
\be
I =\int_{\mathbb{R}^{2,1}} \Big[ {1 \over 4g^2} F_{mn} {1 \over \sqrt{-\partial^2}} F_{mn}
+ {1 \over 1 + \xi} (\partial_m A^m) {1 \over \sqrt{-\partial^2}} (\partial_m A^m)
+ \overline{\psi} (i \partial \hskip-0.24cm / + A \hskip-0.24cm /) \psi \Big]. \,\,\,\,
\ee
The coupling parameter $g$ descends from (3+1)-dimensional super Yang-Mills theory, and hence is dimensionless. Therefore, the (2+1)-dimensional nonlocal conformal field theory is characterized by
gauge group U(1), gauge coupling parameter $g$ and Dirac fermion flavors $N_f$.

Recall that mass term of two-component fermion is parity symmetry violating, whereas mass term of four-component Dirac fermion is invariant under generalized parity. Therefore, as the parameters
$g$ and $N_f$ are varied, it could be that the fermions develop a mass gap dynamically. This is an
indication that the conformal field theory develops a quantum phase transition. In the context of (2+1)-dimensional gauge theories, dynamical mass generation and chiral symmetry breaking therein were
studied extensively. By dimensional analysis, the mass gap generated is proportional to square of the gauge coupling parameter (which has the mass dimension in (2+1) dimensions). In the present context, however, $g$ is dimensionless. I thus claim that fermion mass gap, if generated by strong gauge dynamics, ought to be proportional to the ultraviolet cutoff $\Lambda$.

The conformal invariance SO(3,2) $\subset$ SO(4,2) poses an
interesting question to the traditional view on dynamical generation
of fermion masses. From the dynamical mass generation viewpoint, the
whole reason why 3-dimensional gauge theory with {\sl local}
interactions is interesting is because the gauge coupling parameter
is dimensionful. Thus, if the fermion mass is generated dynamically,
it ought to be proportional to the gauge coupling parameter:
\be m_{\rm dyn} \propto \Big[g_3^2 \Big]_{\rm local}. \ee
From the outset, there is no conformal invariance SO(3,2) and it is
far easier for the fermion to acquire dynamically generated mass. In
particular, since the theory is super-renormalizable, there is no
need to introduce ultraviolet cutoff $\Lambda_{\rm UV}$. As such,
the dynamically generated mass is independent of $\Lambda_{\rm UV}$.

Here, it is quite different. The (classical) Lagrangian is invariant
under SO(3,2) conformal transformation. The effective field theory
is renormalizable, as can be seen from the power-counting:
the gauge field propagator is given by $D_{mn} \propto 1/p$ instead
of $1/p^2$. Therefore, if the fermion mass were generated
dynamically, it should be proportional to the ultraviolet cutoff
$\Lambda_{\rm UV}$ since this is the only dimensionful scale in the
theory. This argument may "explain" why the D-brane setup shows
D7-brane deformation is typically of order the ultraviolet cutoff of the
${\cal N}=4$ super Yang-Mills theory. I see that the present string theory
setup realizes so-called "conformal quantum phase transition" put
forward by Miransky and Yamawaki and studied by many others.

Indeed, our string theory setup features characteristics of the
{\sl conformal phase transition}. They are: \hfill\break
$\bullet$ While the standard phase transition is controlled by
strength of relevant operators (e.g. mass-squared term in
Landau-Ginzburg description of phase transition), the conformal
phase transition is controlled by strength of marginal operators. In
the string theory setup, underlying SO(4,2) invariance and manifest
SO(3,2) invariance forbids contribution of any relevant operators.
\hfill\break
$\bullet$ An order-parameter of the phase transition is given by the
mass gap. I have shown in field theory computation that the mass gap
changes from zero to nonzero value as the (conformal) gauge coupling
is varied continuously. It implies that the low-energy spectrum
changes nonanalytically across the critical line. This is clearly
the characteristics of so-called quantum criticality and quantum
phase transition. At weak coupling, the spectrum includes
(2+1)-dimensional defect fermions interacting with (3+1)-dimensional
gauge fields.
\hfill\break
$\bullet$ I showed that dynamically generated mass is proportional to
$\Lambda$, and hence ultraviolet divergent in the phase of dynamical
mass generation. A function of $\lambda$ multiplies the divergence,
but it cannot be renormalized away in any conventional manner. The
introduction of the ultraviolet cutoff $\Lambda$ is indispensable
for defining the SD gap equation. This is the source of quantum
breaking of the conformal invariance.

I can determine the dynamical mass gap by solving Schwinger-Dyson equation via rainbow approximation and
nonlocal gauge. I only quote the final result:
\be
\Sigma (p^2) \simeq \sqrt{M^3 \over p} \sin \Big( {\kappa \over 2} \log {p \over M} \Big)
\ee
where $M \equiv \Sigma(M^2)$ and $\kappa^2 = (8g^2/3\pi^2)/(1 + g^2 N_f /8) - 1>0$. I bring out interesting
features of the result. First, oscillatory behavior has to do with supercritical instability above critical line defined by $\kappa = 0$ in $(N_f, g)$ parameter space. Second, mass gap is generated at the infrared as back-reaction to the supercritical instability. Third, across the critical line, generation of fermion mass gap implies quantum phase transition between metal and insulator phases.

I can also determine the dynamical mass gap by solving Bethe-Salpeter kernel in two-particle channel. Again, the SO(3,2) conformal invariance plays an important role. The result is in agreement with the above Schwinger-Dyson approach and with onset of supercritical instability phenomena. The massless spectrum includes Goldstone bosons on Grassmannian $U(2N_f)/U(N_f) \times U(N_f)$.

I can also explain pattern of flavor and parity symmetry breaking in terms of D-brane complex. The fermion
flavor is rotated by U($2N_f$) transformation, while the generalized parity transformation maps $(x, y, t) \rightarrow (x, - y, t)$ and pairwise flavor exchange. According to the Vafa-Witten theorem, the flavor symmetry group is broken as U($2N_f) \rightarrow$ U($N_f) \times$ U($N_f$), for which the parity symmetry is unbroken. I can explain these results intuitively terms of the D-brane complex. Recall that the complex has codimension-6, so D7-branes are repelled by the D3-branes. That is, along the overall $x^9$-direction,  D7-branes are pushed away either to the left or to the right of D3-branes. On the other hand, these D7-branes source Ramond-Ramond $G_2 = \star^{10} G_9$ flux along $x^9$ direction. Much like in two-dimensional Schwinger model, the flux jumps across each D7-branes. It is easy to see that the flux energy is minimized when $2N_f$ D7-branes are split to $N_f$ to the left of D3-branes and to $N_f$ to the right. Evidently, the
configuration is $x^9 \rightarrow - x^9$ reflection symmetric, and also explains why the parity symmetry is
preserved.

I now take the limit $N_c \rightarrow \infty$ while holding $\lambda = g_s N_c$ fixed, and focus on low-energy limit. This amounts to replacing the D3-branes in the complex by AdS$_5 \times \mathbb{S}^5$ near-horizon geometry with radius of curvature $L = (4 \pi g_s N_c)^{1/4}$. The D7-branes are then located on calibrated eight-dimensional hypersurface and denotes $2N_f$ gapless fermions in quenched approximation. The calibrated D7-brane also couples to the gapless fermions. Denoting location of $2N_f$ D7-branes along $x^9$-direction as $X_9$, the fermions couple to $X_9$ via
\be
I_{\rm mass} = \int_{\mathbb{R}^{2,1}} \sum_{a,b=1}^{2N_f} \overline{\psi}_a X_9^{ab} \psi_b \, .
\ee
It shows that profile of $X^9$ in AdS$_5 \times \mathbb{S}^5$ provides holographic realization of dynamical mass generation and flavor / parity symmetry breaking pattern. Therefore, to understand strong coupling counterpart of the Schwinger-Dyson analysis, it suffices to study $X^9$ profile of the D7-branes as a function of radial distance $r$ transverse to the center of AdS$_5$, where black D3-branes are situated. By solving Dirac-Born-Infeld action of D7-brane, I found that the solution behaves as
\be
X^9(r) \simeq \sqrt{L^3\over r} \sin \Big( {\kappa_s \over 2} \log {r \over L} \Big).
\ee
at $r \gg L$ and constant at $r \ll L$. Again, the oscillatory behavior is an indicative of violation of
Breitenlohner-Freedman bound: excitation on D7-brane is not confined within D7-branes but can leak to AdS$_5 \times \mathbb{S}^5$ near-horizon geometry. This fits well with the weak coupling analysis in that the dynamically generated mass scale is set by ultraviolet cutoff. If I extend the bulk spacetime to full D3-brane geometry, the instability is cutoff and $X^9$ falls off as $\sim 1/r^3$ sourced by a point source on D7-brane.

Pattern of flavor and parity symmetry breaking can be understood by considering energy density of Ramond-Ramond 1-form $G_1 = \star^{10} G_9$ sourced by D7 branes. I found that minimum of the energy density is when U$(2 N_f) \rightarrow$ U($N_f) \times$ U$(N_f$), at which the generalized parity symmetry is also
unbroken. For partially quenched extension, I also found that the pattern persists, U$(2N_f \vert 2 N_g)
\rightarrow {\rm U}(N_f \vert N_g) \times {\rm U}(N_f \vert N_g)$.

\section{Holography of Fermi Surface}
Here, I present another setup realizing Fermi surfaces \footnote{This is based on a work in collaboration with D. Bak \cite{bakrey}.}. Consider $N_c$ D3-branes in the Coulomb branch, distributed and rotating on a compact domain in transverse $\mathbb{R}^6$-space. Optionally, I can also introduce $N_f$ D7-branes
that sweep entire AdS$_5$-space and three cycle inside $\mathbb{S}^5$. In this case, the gauge field in the foregoing discussions is interpretable as U(1) subgroup of flavor symmetry group of the fermions. Additionally, the ratio between the gravitational and the gauge couplings is suppressed further by $N_f/N_c$. Otherwise, all the physics are identical.  At weak coupling, dynamics of the setup is described by ${\cal N}=4$ super Yang-Mills theory. In addition to gauge bosons, it contains four Majorana fermions and six Hermitian scalars, transforming under SU(4)$_R$ symmetry in ${\bf 4}$ and ${\bf 6}$ representations, respectively. At a generic point in Coulomb branch, SU(4)$_R$ symmetry is broken to three commuting U(1) groups. Here, I will consider the simplest situation of turning on diagonal
combination of the three U(1) groups \cite{cveticgubser}.

At large $N_c$ and $\lambda$, the setup is described most appropriately by five-dimensional anti-de Sitter
charged black hole:
\be
&& \dd s^2 = {1 \over r^2} [ - f(r) \dd t^2 + \dd {\bf x}^2 + f^{-1}(r) \dd r^2] \qquad \mbox{where}
\qquad f(r) = 1 - ar^4 + {1 \over 12} q^2 r^6 \nonumber \\
&& F_{rt} = q r \, .
\ee
As the setup carries electric charges only, Chern-Simons cubic coupling of U(1) gauge fields does not
play a role.
The metric function $f(r)$ has the minimum at radius $r_m= \sqrt{8a/q^2}$, so the black hole exists provided $f(r_m) \le 0$. This puts the so-called singularity-free condition:
\be
64 a^3 \ge 3 q^4.
\ee
The black hole horizon $r_H$ is defined by the smallest root of the equation $f(r_H) = 0$, and determines the Hawking temperature
\be
T_H = {1 \over 4 \pi} \vert f'(r_H)\vert = {a r_H^3 \over \pi} \Big( 1 - {q^2 r_H^2 \over 8a} \Big)
\ee
From the configuration, I can determine energy density $\epsilon$, pressure $p$, charge density $\rho$
 and chemical potential $\mu$. Using the relation $1/(16 \pi G_5) = N_c^2 / 8 \pi^2$, they are
\be
\epsilon &=& {N_c^2 \over 8\pi^2} 3 a \nonumber \\
p &=& {\epsilon \over 3} \nonumber \\
\rho &=& {N_c^2 \over 8 \pi^2} q \nonumber \\
\mu &=& {1 \over 2} r_H^2 q.
\ee
In terms of these quantities, the singularity-free condition reads
\be
\Big({9 \over 8}\Big)^2 \rho^4 \le \gamma \epsilon^3 \qquad \mbox{where} \qquad
\gamma \equiv {N_c^2 \over 8 \pi^2}. \label{gravityside}
\ee

In viewpoint of AdS/CFT correspondence, what is this charged black hole dual to? I claim that the setup describes Fermi surface. The existence of Fermi surface is not necessarily in contradiction with strong
coupling regime of the dual ${\cal N}=4$ super Yang-Mills theory at finite R-charge density. To substantiate
the claim, imagine turning off interactions and take noninteracting limit of fermions at zero temperature. Total number and total energy of the gas per single species are
\be
 N &=& \Big({L \over 2 \pi} \Big)^3 4 \pi \int_0^{p_F} \dd p \, p^2 = {V \over 8 \pi^2} \cdot {4 \over 3} p_F^3
\nonumber \\
E &=& \Big({L \over 2 \pi}\Big)^3 4 \pi \int_0^{p_F} \dd p \, p^3 = {V \over 8 \pi^2} p_F^4.
\ee
Hence, if I denote numerical factor that goes with the total degrees of freedom as $N_c^2 \sigma$, charge and energy density are
\be
\rho_F = \gamma \sigma_B {4 \over 3} p_F^3 \qquad \mbox{and} \qquad
\epsilon_F = \gamma \sigma_E p_F^4.
\ee
If fermions are intercting, the energy density should be higher. So, I have
\be
\Big({\sigma_E^3 \over 4 \sigma_B^4} \Big) \Big({9 \over 8}\Big)^2 \rho_F^4 \le \gamma \epsilon^3_F \, .
\label{freefermionside} \ee
Comparison between (\ref{freefermionside}) and (\ref{gravityside}) shows that the conformal field theory and the dual supergravity results agree if $\sigma_B = \sigma_E = 1/4$. Though these numbers are not {\sl ab initio} calculable from strong coupled ${\cal N}=4$ super Yang-Mills theory, the fact that they are predicted to be constant of order unity lenders strong support to my proposal.

To substantiate the proposal, I now consider turning on a finite but small temperature to the system. Then, my proposal asserts that non-extremal charged AdS$_5$ black hole with Hawking temperature $T_H$ is dual to (3+1)-dimensional relativistic Fermi surface at temperature $T = T_H$. To see this, I perturb the black hole from the extremal value $r_H = r_c, a = a_c$ by tuning parameters $a, q$:
\be
a = a_c[1 + b \kappa^2 + {\cal O}(\kappa^3)], \qquad
r_H  = r_c[1 - \kappa]
\ee
while holding the charge density fixed. I find the coefficient $b=4$ from the horizon condition $f(r_H) = 0$. I also find the Hawking temperature
\be
T_H = {q^2 r_c^2 \over 4 \pi} \kappa = \Big({2 \over \pi} a_c^{1/4} 3^{3/4} \Big) \kappa = T.
\ee
Solving for $a$ perturbatively, from the relation ${\cal E} = 3 \gamma a$, I obtain
\be
{\cal E} = {\cal E}_c \Big[ 1 + {\pi^2 \over 3} \sqrt{\gamma \over {\cal E}_c} T^2 + {\cal O}(T^3)\Big].
\ee
It would be interesting to compare this result with noninteracting fermion gas at fixed fermion charge. An elementary computation at low temperature indicates that
\be
{\cal E} = {\cal E}_0 \Big[ 1 + {\pi^2 \over 3} \sqrt{4 \sigma_E \gamma \over {\cal E}_0} T^2 + {\cal O}(T^3) \Big].
\ee
Again, I find perfect agreement with the black hole result provided I fix the numerical factor $\sigma_E = 1/4$.

For those who has not converted yet, I now provide further evidence for my proposal that charged black hole describes Fermi surface. I shall now study dynamical perturbation of the charged black hole and compare with
out-of-equilibrium dynamics of Fermi gas. The simplest sort of the latter is the dynamic structure factor \cite{nozierespines}
\be
S(\omega, {\bf k}) = \int \dd^4 x \langle J_0(x) J_0 (0) \rangle e^{- i k \cdot x}
\ee
This is related to the density response function $G_{mn}$ by
\be
G_{00}(\omega, {\bf k}) = \int {\dd \omega' \over 2 \pi} {1 \over \omega - \omega' +i \epsilon}
[S(\omega, {\bf k}) - S(- \omega, - {\bf x})].
\ee
For non-interacting, spinless, relativistic fermion with finite density,
\be
S(\omega, {\bf k}) = \int {\dd^3 {\bf q} \over (2 \pi)^2} \delta( \omega - (|{\bf k} + {\bf q}| - |{\bf k}|)) \theta (|{\bf k} +{\bf q}| - p_F) \theta (p_F - |{\bf k}|).
\ee
For $|{\bf k}| \ll p_F$, this is easily evaluated as
\be
S(\omega, {\bf k}) = {p_F^2 \over 2 \pi} {\omega \over |{\bf k}|} \theta(|{\bf k}| \omega - \omega^2)
\ee
So, for $\omega \gg |{\bf k}|$, I find that
\be
G_{00}(\omega, {\bf k}) = N_c^2 \sigma_B {p_F^2 \over 6 \pi^2} {|{\bf k}|^2 \over \omega^2} = {3 \over 4} {\sigma_E \over \sigma_B} {\rho^2 \over {\cal E}} {|{\bf k}|^2 \over \omega^2},
\label{freefermiongoo} \ee
where $N_c^2 \sigma_B$ is identified with the number of fermions.

I shall now compute the corresponding observable from the black hole side. As mentioned, I need to compute
linearized fluctuation of the black hole background. The requisite boundary conditions are that the fluctuation is incoming and outgoing at radial infinity $r \rightarrow 0$ and purely incoming at black hole
horizon $r = r_H$. Along $\mathbb{R}^{3,1}$, the fluctuations are travelling waves proportional to $e^{- i \omega t + i {\bf k} \cdot {\bf x}}$. I shall take ${\bf k} = k \hat{\bf e}_3$ and decompose fluctuation of the metric $h_{mn}(r, x)$ and the gauge field $A_m (r, x)$ components into longitudinal and transverse components with respect to $\hat{\bf z}$. To compare with (\ref{freefermiongoo}), I need to study vector modes. Choosing the gauge $h_{r1}=h_{r2}=0$ and vector polarization along $\hat{e}_1$ direction, I find that the on-shell generating functional for the fluctuations $a_0 = g^{11} h_{10}, a_3 = g^{11} h_{13}, A=A_1$
becomes
\be
I_{\rm vector} = \gamma \int \dd^4 x \Big[ - {1 \over y} a_0' a_0 + {h \over y} a_3' a_3 + h A' A
+ {q \over 2} A a_0 - {a \over 2} a_0^2 - {a \over 2} a_3^2 \Big]_{y=0}
\ee
where a change of variable is made to $r = y^2$. I also need to add the extrinsic curvature term and local counterterm for holographic renormalization. After some computations, I find that
\be
G_{00} (\omega, {\bf k}) = K {i \omega \over i \omega - D |{\bf k}|^2} {|{\bf k}|^2 \over \omega^2}
\qquad \mbox{where}\qquad K = {N_c^2 \over 8 \pi^2} {q^2 \over 4a} = {3 \over 4} {\rho^2 \over {\cal E}}.
\ee
Comparing now this result with the result for non-interacting, colored, relativistic fermion in
(\ref{freefermiongoo}), I once again find perfect agreement provided I choose $\sigma_B = \sigma_E = 1/4$!
I also confirmed that, by repeating the computations for non-extremal black hole, the agreement also extends to finite temperature.

I also studied the tensor mode $g^{11} h_{11}$. It satisfies massless Laplace equation in charged black hole background. By solving the fluctuation equation with the incoming boundary condition at the horizon, I extracted the retarded Green function
\be
G_{1212}(\omega, {\bf k}) &=& - i \int \dd^4 x \theta(t) e^{- i k \cdot x} \langle [T_{12}(x), T_{12}(0)] \rangle  \nonumber \\
&=& \gamma {i \omega \over r_H^3} + {\cal O}(\omega^2, {\bf k}^2).
\ee
From this follows the shear viscosity
\be
\eta = \lim_{\omega \rightarrow 0} \lim_{k\rightarrow 0} {1 \over \omega} G_{1212}(\omega, {\bf k}) = {N_c^2 \over 8 \pi^2} {1 \over r_H^3}.
\ee
Comparing this to the entropy density
\be
s = {S \over V_3} = {N_c^2 \over 2 \pi} {1 \over r_H^3}
\ee
I find that the system saturates viscosity-to-entropy ratio
\be
{\eta \over s} = {1 \over 4 \pi}.
\ee
This indicates that, even though thermodynamics and transport exhibits Landau's fermi liquid behavior, the system is intrinsically strongly interacting.

I will present one more check-point. The stopping power  -- rate of energy loss per unit distance -- of a heavy particle moving in velocity $V$ in fermi gas is defined by \cite{nozierespines}
\be
{\dd E \over \dd x} = {1 \over V} {\dd E \over \dd t} = {1 \over V^2} \Gamma.
\ee
It is known in neutral Fermi liquid that $\Gamma$ depends on the interaction potential and integrate dynamical structure factor:
\be
\Gamma = \int_0^\infty \dd k k |V_{\rm k}|^2 \int_0^\infty {\dd \omega \over 2 \pi} S(\omega, {\bf k}).
\ee
This observable is also extractable from charged black hole. Consider a probe heavy fermion. It is described by a fundamental string dangling from the boundary to the interior toward the black hole. Since the charged
black hole background breaks Lorentz invariance, the string will be dragged near the black hole horizon and exerts a force. This is computed straightforwardly by solving Nambu-Goto equation of the string. The result is
\be
F_{\rm drag} = {\dd E \over \dd x} = T_F {V \over r_B^2}
\ee
where $r_B$ is fixed by the energy condition $V^2= f(r_B) = 1 - a r_B^4 + {q^2 \over 12} r_B^6$. I found
that $1/r_B^2 = \sqrt{a}/\sqrt{1 - V^2} \kappa$, where $\kappa^2 = 1 - q^2 r_B^2/12a$. Therefore, I extract
the drag force
\be
F_{\rm drag} = T_F {V \over \sqrt{1 - V^2}} \sqrt{a} \kappa(V).
\ee
From the definition, one can show that $1/3 \le \kappa^2 \le 1$ and that ultra-relativistic limit asymptotics $\kappa(V=1)=1$. Therefore, I confirm that, even at zero temperature, fast moving fermion experiences the drag force due to interaction with the Fermi liquid media. Once again, I observe that this qualitative feature fits well with Landau's fermi liquid theory.

There is one important caveat to the proposal. While charge (or chemical potential) and temperature
dependent part of equilibrium thermodynamics and out-of-equilibrium dynamics of the AdS$_5$ charged black hole is tantalizingly similar to the $(3+1)$-dimensional non-interacting Fermi gas, it cannot be all that
same. For one thing, the extremal charged black hole has nonzero entropy while the Fermi gas at zero temperature does not. For another, the zero temperature $G_{00}(\omega, {\bf k})$ correlator exhibits diffusive behavior instead of ballistic behavior. This means that the black hole does not precisely correspond to Fermi gas. A viable alternative is that the dual is not pure Fermi gas but disordered Fermi gas --- Fermi gas interacting with some sort of homogeneous disorder. The zero temperature entropy is then interpretable as entropy of the disorder if the disorder is quantum-mechanically strongly entangled. The diffusive behavior is also interpretable as a consequence of interaction between the Fermi gas and the
disorder. As of now, I do not have any smoking gun evidence that this is the right interpretation. Clarifying whether this is correct or some other picture is so would be a very interesting problem left for future
investigation.

\section{Prospects}
I hope in this talk to have conveyed you that salient features of string theory can be encoded into high mobility electron gas inside thin semiconductors. This is novel and exciting. By making the connection more
precise, I believe we shall even be able to test string theory experimentally with added advantage that it is cost effective. Even apart from this prospects, making contact the string theory with seemingly unrelated physical phenomena is always exciting and would bring yet another layer of surprises and wonders.

\section*{Acknowledgements}
For the topics covered in this talk, I acknowledge Dongsu Bak, David J. Gross, David Kutasov, Juan Maldacena, Volodya Miransky and Subir Sachdev for many enlightening discussions. I am grateful to organizers of the conference for providing stimulating atmosphere and warm hospitality.

%

\end{document}